\documentclass[12pt]{article}
\usepackage{amsfonts}

\catcode`\@=11

\newcount\hour
\newcount\minute
\newtoks\amorpm \hour=\time\divide\hour by 60\minute
=\time{\multiply\hour by 60 \global\advance\minute by-\hour}
\edef\standardtime{{\ifnum\hour<12 \global\amorpm={am}%
        \else\global\amorpm={pm}\advance\hour by-12 \fi
        \ifnum\hour=0 \hour=12 \fi
        \number\hour:\ifnum\minute<10
        0\fi\number\minute\the\amorpm}}
\edef\militarytime{\number\hour:\ifnum\minute<10
0\fi\number\minute}

\def\draftlabel#1{{\@bsphack\if@filesw {\let\thepage\relax
   \xdef\@gtempa{\write\@auxout{\string
      \newlabel{#1}{{\@currentlabel}{\thepage}}}}}\@gtempa
   \if@nobreak \ifvmode\nobreak\fi\fi\fi\@esphack}
        \gdef\@eqnlabel{#1}}
\def\@eqnlabel{}
\def\@vacuum{}
\def\marginnote#1{}
\def\draftmarginnote#1{\marginpar{\raggedright\scriptsize\tt#1}}
\def\draft{
        \pagestyle{plain}
        \overfullrule=2pt
        \oddsidemargin -.5truein
        \def\@oddhead{\sl \phantom{\today\quad\militarytime} \hfil
        \smash{\Large\sl DRAFT} \hfil \today\quad\militarytime}
        \let\@evenhead\@oddhead
        \let\label=\draftlabel
        \let\marginnote=\draftmarginnote
        \def\ps@empty{\let\@mkboth\@gobbletwo
        \def\@oddfoot{\hfil \smash{\Large\sl DRAFT} \hfil}
        \let\@evenfoot\@oddhead}
        \def\@eqnnum{(\theequation)\rlap{\kern\marginparsep\tt\@eqnlabel}%
        \global\let\@eqnlabel\@vacuum}  }

\newcommand{\rf}[1]{(\ref{#1})}
\renewcommand{\theequation}{\thesection.\arabic{equation}}
\renewcommand{\thefootnote}{\fnsymbol{footnote}}
\newcommand{\newsection}{    
\setcounter{equation}{0}\section}
\def\appendix#1{\addtocounter{section}{1}\setcounter{equation}{0}
\renewcommand{\thesection}{\Alph{section}}
\section*{Appendix \thesection\protect\indent \parbox[t]{11.15cm}{#1}}
\addcontentsline{toc}{section}{Appendix \thesection\ \ \ #1}}

\def\AA{A}

\def\MTIL{{\widetilde M}}

\def\half{\frac{1}{2}}

\def\bfP{{\bf P}}
\def\bfJ{{\bf J}}
\def\bfM{{\bf M}}

\newcommand{\Co}{\mathbb{C}}

\def\be{\begin{equation}}
\def\ee{\end{equation}}
\def\beq{\begin{eqnarray}}
\def\eeq{\end{eqnarray}}

\begin{document}


\begin{flushright}
FIAN/TD/11-04 \\
hep-th/0412311
\end{flushright}

\vspace{1cm}

\begin{center}

{\Large \bf Mixed-symmetry massive fields in AdS(5)}

\vspace{2.5cm}

R.R. Metsaev\footnote{ E-mail: metsaev@lpi.ru }

\vspace{1cm}

{\it Department of Theoretical Physics, P.N. Lebedev Physical
Institute, \\ Leninsky prospect 53,  Moscow 119991, Russia }

\vspace{3.5cm}

{\bf Abstract}

\end{center}

Free mixed-symmetry arbitrary spin massive bosonic and fermionic
fields propagating in AdS(5) are investigated. Using the
light-cone formulation of relativistic dynamics, we study bosonic
and fermionic fields on an equal footing. Light-cone gauge actions
for such fields are constructed. Various limits of the actions are
discussed.

\newpage
\renewcommand{\thefootnote}{\arabic{footnote}}
\setcounter{footnote}{0}

\section{Introduction}

The study of higher spin {\it massive} fields theories in $AdS$
space-time is motivated by the conjectured duality of the
conformal ${\cal N}=4$ SYM theory and the theory of type $IIB$
superstring in $AdS_5 \times S^5$ background \cite{malda}. Recent
interesting discussions of various aspects of this theme may be
found, e.g., in
\cite{Tseytlin:2004xa,Arutyunov:2004vx,Bianchi:2004xi,Belitsky:2004cz}.
As is well known, quantization of GS superstring propagating in
flat space is straightforward only in the light-cone gauge. It is
the light-cone gauge that removes unphysical degrees of freedom
explicitly and reduces the action to the quadratic form in string
coordinates. The light-cone gauge in string theory implies the
corresponding light-cone formulation for target space fields. In
the case of strings in $AdS$ background, this suggests that we
should study a light-cone form dynamics of {\it target space
fields} propagating in $AdS$ space-time. It is expected that $AdS$
massive fields form the spectrum of states of $AdS$ strings.
Therefore, understanding the light-cone description of $AdS$
massive target space fields might be helpful in discussion of
various aspects of $AdS$ string dynamics. This is what we do in
this paper.

Let us first formulate the main problem we solve in this paper.
Fields propagating in $AdS_5$ space are associated with
positive-energy unitary lowest weight representations of the
$SO(4,2)$ group. A positive-energy lowest weight irreducible
representation of the $SO(4,2)$ group, denoted as $D(E_0,{\bf
h})$, is defined by $E_0$, the lowest eigenvalue of the energy
operator, and by ${\bf h}=(h_1, h_2)$,  which is the highest
weight of the unitary representation of the $SO(4)$ group. The
highest weights $h_i$ are integers and half-integers for bosonic
and fermionic fields respectively and satisfy the standard
restriction
\be h_1 \geq |h_2|\,.\ee
The fields with ${\bf h}=(0,0)$ and ${\bf h}=(1/2,\pm1/2)$
correspond to the respective scalar bosonic and spin one-half
fermionic fields. The actions for such fields are well known.
$E_0$ and ${\bf h}$ associated with the remaining fields in
$AdS_5$ are given by
\be\label{bose01x} \hspace{1.1cm} E_0 > h_1 + 1\,, \qquad h_1 =
|h_2|>1/2\,, \ee
\be\label{bose01xx}  E_0 = h_1 + 2\,, \qquad h_1 > |h_2|\,, \ee
\be\label{bose01}  E_0 > h_1 + 2\,, \qquad h_1 > |h_2|\,. \ee
The fields with $E_0$,${\bf h}$ in \rf{bose01xx} and \rf{bose01x}
are referred to as massless and self-dual massive\footnote{
We recall that the representations with $E_0=h_1+1$, $h_1=|h_2|$
do not admit field theoretical realization in $AdS_5$.}
fields respectively. The fields with $E_0$, ${\bf h}$ in
\rf{bose01} are referred to as massive fields and these fields can
be divided into two groups
\beq \label{bose01x1}  && E_0 > h_1 + 2\,, \qquad h_1 >
|h_2|\,,\qquad  h_2=0 ,\pm 1/2\,,
\\
\label{bose01x2} && E_0 > h_1 + 2\,, \qquad h_1 > |h_2|\,, \qquad
|h_2| >1/2\,. \eeq
The massive fields in \rf{bose01x1} are referred to as totally
symmetric bosonic ($h_2=0$) and fermionic ($h_2= \pm 1/2$) fields
respectively, while the massive fields in \rf{bose01x2} are
referred to as mixed-symmetry fields\footnote{
We note that the case ${\bf h}=(1,0)$ corresponds to spin one
massive field, the case ${\bf h}=(2,0)$ is the massive spin two
field. The labels $h_i$ are the standard Gelfand--Zeitlin labels.
They are related to Dynkin labels $h_i^D$ by the formula
$(h^D_1,h^D_2)= (h_1-h_2, h_1 + h_2)$.}.
In manifestly Lorentz covariant formulation the bosonic(fermionic)
totally symmetric and mixed-symmetry massive representation are
described by a set of the tensor(tensor-spinor) fields whose
$SO(4,1)$ space-time tensor indices have the structure of the
Young tableaux with one and two rows respectively. Lorentz
covariant actions for the {\it bosonic totally symmetric massive}
fields in $AdS_d$ space were found in \cite{zin}\footnote{
Massive self-dual spin fields in $AdS_3$ were investigated in
\cite{rrm01088}. Spin two massive fields were studied in
\cite{pol} (see also \cite{Buchbinder:1999ar,Buchbinder:2000fy}).
A discussion of massive totally symmetric fields in $(A)dS_d$,
$d\geq 4$, may be found in \cite{des}. Description of massive
fields on constant curvature spaces corresponding to  two-column
Young tableaux irreducible representations of the general linear
group may be found in \cite{deMedeiros:2003px}.}.
Light-cone actions for both the {\it bosonic and fermionic totally
symmetric massive} fields in $AdS_d$ with arbitrary $E_0$ and
${\bf h}$ were obtained in \cite{Metsaev:2003cu}. Light-cone
actions for $AdS_5$ {\it mixed-symmetry massless} \rf{bose01xx},
$|h_2|>1/2$, {\it and self-dual massive} fields \rf{bose01x} were
found in \cite{rrm02226}\footnote{
Lorentz covariant actions for the mixed-symmetry massless fields
in $AdS_5$ were obtained recently in \cite{Alkalaev:2005kt}(see
also \cite{Brink:2000ag},\cite{Alkalaev:2005kw}).}. {\it
Mixed-symmetry massive} $AdS_5$ fields \rf{bose01x2} (bosonic and
fermionic ones) have not been described at the field theoretical
level so far\footnote{
Study of unitary representations of the $so(4,2)$ algebra may be
found, e.g., in \cite{Mack:1975je}. Group theoretical description
of various massive representation of the $so(4,2)$ algebra via the
oscillator method \cite{Bars:1982ep} may be found, e.g., in
\cite{gunmin}. Lorentz covariant equations of motion for such
fields with special values of $E_0$ were discussed in \cite{sez2}.
Lorentz covariant actions for mixed-symmetry massive $AdS_d$
fields corresponding to particular values of ${\bf h}$ were
constructed in \cite{Zinoviev:2002ye}.}.
In this paper, we develop a light-cone gauge formulation for such
fields at the action level. In manifestly Lorentz covariant
formulation these mixed-symmetry massive fields correspond to
Young tableaux with two rows\footnote{
Bosonic mixed-symmetry fields correspond  to Young tableaux with
row lengths $h_1$ and $|h_2|$, while fermionic fields correspond
to Young tableaux with row lengths $h_1-\frac{1}{2}$ and $|h_2|-
\frac{1}{2}$.}
$$ \label{dia}
\begin{picture}(130,45)
\put(30,34){$h_1$} {\linethickness{.500mm}
\put(00,20){\line(1,0){140}}%
\put(00,30){\line(1,0){140}}%
\put(00,10){\line(1,0){70}}%
\put(00,10){\line(0,1){20}}%
\put(140,20){\line(0,1){10}}%
\put(70,10){\line(0,1){10}}}%
\put(10,10.0){\line(0,1){20}} \put(20,10.0){\line(0,1){20}}
\put(30,10.0){\line(0,1){20}} \put(40,10.0){\line(0,1){20}}
\put(50,10.0){\line(0,1){20}} \put(60,10.0){\line(0,1){20}}
\put(70,20.0){\line(0,1){10}} \put(80,20.0){\line(0,1){10}}
\put(90,20.0){\line(0,1){10}} \put(100,20.0){\line(0,1){10}}
\put(110,20.0){\line(0,1){10}} \put(120,20.0){\line(0,1){10}}
\put(130,20.0){\line(0,1){10}} 
\put(30,-05){$|h_2|$}
\end{picture}
$$

Using a new version \cite{Metsaev:2003cu} of the old light-cone
gauge formalism in $AdS$ space \cite{rrm99217}, we describe both
the bosonic and fermionic fields on an equal footing. Since, by
analogy with flat space, we expect that a quantization of the
Green-Schwarz $AdS$ superstring with a Ramond-Ramond charge will
be straightforward only in the light-cone gauge \cite{mt3}, it
seems that from the stringy perspective of $AdS/CFT$
correspondence the light-cone approach is the fruitful direction
to go\footnote{
Note that sometimes a light-cone gauge formulation turns out to be
a good starting point for deriving a Lorentz covariant
formulation. This is to say that for fields in flat space,
interesting methods were developed \cite{Siegel:1986zi,SIEG} which
admit manifest covariantization of the results obtained in the
light-cone gauge. It is highly likely that these methods can be
generalized to the case of fields in AdS space.}.
An interesting recent discussion of the relationship between the
light-cone gauge and BRST quantizations of superstrings may be
found in \cite{Berkovits:2004tw}.

The content of this paper is as follows. In section 2, we
summarize some relevant facts about light-cone gauge actions and
discuss realization of relativistic symmetries of such actions in
the light-cone approach. Light-cone actions for bosonic and
fermionic fields are constructed in sections 3 and 4 respectively.
In section 5, we discuss various limits of the actions: limits of
massive self-dual fields and massless fields in $AdS_5$ and the
flat space limit. Section 6 suggests directions for future
research. Appendices contain some mathematical details and useful
formulas.

\newsection{Light-cone gauge action and its global symmetries}

In this section we present a new version \cite{Metsaev:2003cu} of
the old light-cone formalism \cite{rrm99217}. Let $\phi(x)$ and
$\psi(x)$ be respective bosonic and fermionic arbitrary spin
fields propagating in $AdS_5$ space. If we collect spin degrees of
freedom in ket-vectors $|\phi\rangle$ and $|\psi\rangle$, then the
respective light-cone gauge actions for the fields $\phi$ and
$\psi$ can be cast into the following `covariant form'
\beq && \label{lcact} S_{l.c.} =\frac{1}{2}\int d^5 x \langle
\phi|\bigl(\Box -\frac{1}{z^2}A\bigr)|\phi\rangle\,, \qquad\quad\
\  \hbox{for bosonic fields},
\\[7pt]
\label{lcactfer} && S_{l.c.}= \int d^5 x\, \langle \psi| \frac{\rm
i}{2\partial^+}(\Box - \frac{1}{z^2}A)|\psi\rangle\,, \qquad
\hbox{for fermionic fields}, \eeq
where a light-cone representation of the flat Lorentz covariant
D'Alembertian is given by (details of the notation may be found in
Appendix A)
\be  \Box = 2\partial^+\partial^- +
\partial_i\partial_i + \partial_z^2\,.
\ee
The operator $A$ does not depend on space-time coordinates and
their derivatives. This operator acts only on spin indices
collected in the ket-vectors $|\phi\rangle$ and $|\psi\rangle$. We
call the operator $A$ the $AdS$ mass operator.

It is instructive to present the $AdS$ mass operator for simplest
cases of the scalar field, ${\bf h}=(0,0)$, and spin one-half
field, ${\bf h}=(1/2,\pm 1/2)$. For a massive scalar field, the
operator $A$ takes the form
\be A=(mR)^2 + \frac{d(d-2)}{4}\,,\qquad d=5\,, \ee
where we show the dependence on the dimension $d$ of $AdS$ space
explicitly, while for the case of a spin one-half Dirac field
corresponding to ${\bf h}=(1/2,\pm 1/2)$,  the operator $A$ takes
the form
\be\label{Afer} A = (Rm)^2 + Rm \gamma^z \ee
that is valid for an arbitrary dimension of $AdS$ space (see
Appendix B).

We now turn to discussion of the $AdS$ mass operator $A$ for
arbitrary spin massive fields. It turns out that the operator $A$
admits the representation \cite{Metsaev:2003cu}
\be \label{adsope} A  =  2B^z+ M^{zi}M^{zi} +
\frac{1}{2}M^{IJ}M^{IJ} -\langle Q_{AdS}\rangle  + \frac{15}{4}\,,
\ee
where $M^{IJ}$ is a spin operator of the $so(3)$ algebra\footnote{
In the $so(2)$ algebra basis, the operator $M^{IJ}$ splits into
$M^{ij},M^{zi}$, $i,j=1,2$.}:
\be\label{d2comrel} [M^{IJ},M^{KL}]=\delta^{JK}M^{IL} +3 \hbox{
terms}\,.\ee
The operator $M^{IJ}$ is acting only on spin degrees of freedom of
wave function $|\phi\rangle$. In formula \rf{adsope}, $\langle
Q_{AdS}\rangle$ is an eigenvalue of the second order Casimir
operator of the $so(4,2)$ algebra for the representation labelled
by $D(E_0,{\bf h})$:
\be\label{casope1} -\langle Q_{AdS}\rangle =E_0(E_0-4)+ h_1 (h_1
+2) +h_2^2\,,\ee
while $B^z$ is the $z$-component of the $so(3)$ algebra vector
$B^I$ that satisfies the  equation \cite{Metsaev:2003cu}\footnote{
We use the notation $(M^3)^{[I|J]}\equiv
\frac{1}{2}M^{IK}M^{KL}M^{LJ} -(I\leftrightarrow J)$, $M^2\equiv
M^{IJ}M^{IJ}$. We note that for the spin operator of the $so(3)$
algebra, one has the relation $(M^3)^{[I|J]} =
-\frac{1}{2}M^2M^{IJ}$.}
\be\label{basequ0} [B^I,B^J] +(M^3)^{[I|J]} + (\langle
Q_{AdS}\rangle -\frac{1}{2}M^2 -  4 )M^{IJ} = 0\,. \ee
As noted, the operator $B^I$ transforms in the vector
representation of the $so(3)$ algebra,
\be\label{bId2tra}
[B^I,M^{JK}]=\delta^{IJ}B^K-\delta^{IK}B^J\,.\ee
This operator should be hermitian with respect to an appropriate
scalar product that will be discussed below,
\be \label{BIunicon} B^{I \dagger} =B^I\,.\ee

We now turn to discussion of global $so(4,2)$ symmetries of the
light-cone gauge actions of arbitrary spin fields. The choice of
the light-cone gauge spoils the manifest global symmetries, and in
order to demonstrate that these global invariances are still
present, one needs to find the Noether charges which generate
them\footnote{
These charges play  a crucial role  in formulating interaction
vertices in field theory. Application of Noether charges in
formulating superstring field theories may be found in
\cite{gsb}.}.
Noether charges (or generators) can be split into kinematical and
dynamical generators. For $x^+=0$, the kinematical generators are
quadratic in the physical field $|\phi\rangle$, while the
dynamical generators receive corrections in the interacting
theory. In this paper, we deal with free fields. Let us first
consider bosonic fields. At the quadratic level both kinematical
and dynamical generators have the following standard
representation in terms of the bosonic physical field
\cite{rrm99217}:
\be\label{hatG} \hat{G}=\int d^4
x\langle\partial^+\phi|G|\phi\rangle\,,\qquad d^4x\equiv
dx^-\,dz\,d^2x\,. \ee
The representation for the kinematical generators in terms  of
differential operators $G$ acting on the bosonic physical field
$|\phi\rangle$ is given by
\beq \label{pi}&& P^i=\partial^i\,, \qquad  P^+=\partial^+\,,
\\
&& D=x^+ P^- +x^-\partial^++x^I\partial^I + \frac{3}{2}\,,
\\
&& J^{+-}=x^+ P^- -x^-\partial^+\,,
\\
\label{J+i} && J^{+i}=x^+\partial^i-x^i\partial^+\,,
\\
\label{Jij}&& J^{ij} = x^i\partial^j-x^j\partial^i + M^{ij}\,,
\\
&& K^+ = -\frac{1}{2}(2x^+x^-+x^Jx^J)\partial^+ + x^+D\,,
\\
&& K^i = -\frac{1}{2}(2x^+x^-+x^Jx^J)\partial^i +x^i
D+M^{iJ}x^J+M^{i-}x^+\,, \eeq
while the representation for the dynamical generators takes the
form
\beq && P^-=-\frac{\partial^J\partial^J}{2\partial^+}
+\frac{1}{2z^2\partial^+}A\,,
\\
&& J^{-i}=x^-\partial^i-x^i P^-
+M^{-i}\,,\\
\label{km}&& K^-=-\frac{1}{2}(2x^+x^- + x^I x^I) P^- +
x^-D+\frac{1}{\partial^+}x^I\partial^JM^{IJ}
-\frac{x^i}{2z\partial^+}[M^{zi},A] +\frac{1}{\partial^+}B\,,\ \ \
\ \ \ \ \ \ \ \ \ \eeq
where
\be M^{-i} \equiv M^{iJ}\frac{\partial^J}{\partial^+}
-\frac{1}{2z\partial^+}[M^{zi},A]\,,\qquad M^{-i}=-M^{i-}\ee
and the new operator $B$ that enters $K^-$ in \rf{km} admits the
representation
\be \label{bope} B =  B^z + M^{zi}M^{zi}\,. \ee
Making use of the above formulas, one can check that the
light-cone gauge action \rf{lcact} is invariant under the global
symmetries generated by the $so(4,2)$ algebra taken in the form
\be\delta_{\hat{G}} |\phi\rangle = G|\phi\rangle\,.\ee
Generalization of above formulas to the case of fermionic fields
is straightforward. This is to say that at the quadratic level,
both kinematical and dynamical generators have the following
standard representation in terms of the fermionic physical
light-cone field
\be\label{hatGfer} \hat{G}^{ferm}=\int d^4 x\langle \psi|G^{ferm}
|\psi\rangle\,, \ee
where differential operators $G^{ferm}$ are obtainable from those
of bosonic fields \rf{pi}-\rf{km} by making there the following
substitution
\be x^- \rightarrow x^- +\frac{1}{2\partial^+}\,.\ee
In addition to this, in expressions for generators in
\rf{pi}-\rf{km}, we should use the spin operator $M^{IJ}$ suitable
for fermionic fields. Defining equation \rf{basequ0} for the
operator $B^I$ and the representations for the operators $A$, $B$
given in \rf{adsope},\rf{bope} do not change.

The action for fermionic fields \rf{lcactfer} is invariant under
the global relativistic symmetries generated by the $so(4,2)$
algebra:
\be\delta_{\hat{G}^{ferm}} |\psi \rangle =
G^{ferm}|\psi\rangle\,.\ee

To summarize the procedure for finding light-cone description
consists of the following steps:

i) choose the form of a realization of spin degrees of freedom for
the field $|\phi\rangle$ (or $|\psi\rangle$);

ii) fix an appropriate representation for the spin operator
$M^{IJ}$;

iii) find a solution to the defining equations for the operator
$B^I$ in \rf{basequ0}.

Following this procedure, we now discuss bosonic and fermionic
fields in turn.

\newsection{Bosonic fields}

To discuss field theoretical description of a massive $AdS_5$
field, we could use a {\it complex-valued} tensor field $\phi$
that is associated with the weight-${\bf h}$ representation of the
$so(4)$ algebra. We prefer to decompose such field into traceless
totally symmetric complex-valued tensors of the $so(3)$ algebra
$\phi^{I_1\ldots I_{s'}}$, $I=1,2,3$; $s'=|h_2|,|h_2|+1,\ldots,
h_1$:
\be \label{sof00} \phi = \sum_{s'=|h_2| }^{h_1} \oplus\,
\phi^{I_1\ldots I_{s'}}\,. \ee
The number of complex-valued degrees of freedom ($DoF$) of
mixed-symmetry massive field $\phi$ associated with the
weight-${\bf h}$ representation of the $so(4)$ algebra is given by
\be \label{sof01} N_{DoF}^{\Co}(\phi) = (h_1 + 1)^2 -h_2^2\,, \ee
and this number of complex-valued $DoF$ can be presented as
\be \label{sof03} N_{DoF}^{\Co}(\phi)  = \sum_{s'=|h_2|}^{h_1}
(2s'+1)\,. \ee
Formula \rf{sof03} tells us that the tensor field $\phi$ can
indeed be decomposed into totally symmetric tensor fields as
shown in \rf{sof00}.

As usual, to avoid cumbersome tensor expressions, we introduce
creation and annihilation oscillators $\alpha^I$ and
$\bar{\alpha}^I$
\be\label{bososc} [\bar{\alpha}^I,\,\alpha^J]=\delta^{IJ}\,,\qquad
\bar{\alpha}^I|0\rangle =0\,, \ee
and make use of ket-vectors $|\phi_{s'}\rangle$ defined by
\be\label{genfun1} |\phi_{s'}\rangle \equiv \alpha^{I_1}\ldots
\alpha^{I_{s'}} \phi^{I_1\ldots I_{s'}}(x)|0\rangle\,. \ee
The ket-vector $|\phi_{s'}\rangle$ satisfies the algebraic
constraints
\begin{eqnarray}
\label{homcon1}&& (\alpha\bar{\alpha}-s')|\phi_{s'} \rangle
=0\,,\qquad \alpha\bar{\alpha}\equiv \alpha^I\bar\alpha^I\,,
\\
\label{tracon1}&& \bar\alpha^I\bar{\alpha}^I|\phi_{s'}
\rangle=0\,,\qquad \quad \hbox{tracelessness}\,.
\end{eqnarray}
Equation \rf{homcon1} tells us that $|\phi_{s'}\rangle$ is a
monomial of degree $s'$ in the oscillator $\alpha^I$.
Tracelessness of the tensor field $\phi^{I_1\ldots I_{s'}}$ is
reflected in \rf{tracon1}. Thus, in the language of ket-vectors,
the massive field $\phi$ is representable as
\be \label{sof00ket} |\phi\rangle = \sum_{s'=|h_2| }^{h_1}
\oplus\, |\phi_{s'}\rangle \ee
and the scalar product $\langle\varphi||\phi\rangle$ that enters
light-cone action \rf{lcact} is defined to be\footnote{
Because we use complex-valued fields, the bra-vectors
$\langle\phi_{s'}|$ are built in terms of complex conjugate
tensors fields $\phi^{I_1\ldots I_s*}$, where the asterisk is used
to denote complex conjugation.}
\be \label{scaprod01} \langle\varphi||\phi\rangle  \equiv
\sum_{s'=|h_2|}^{h_1} \langle\varphi_{s'}||\phi_{s'}\rangle \,.\ee
The spin operator $M^{IJ}$ of the $so(3)$ algebra for the above
defined fields $|\phi_{s'}\rangle$ then takes the form
\be\label{spiopemij1} M^{IJ} =\alpha^I\bar\alpha^J -
\alpha^J\bar\alpha^I\,.\ee
The action of the operator $B^I$ on the physical fields
$|\phi_{s'}\rangle$ is found to be (details of derivation may be
found in Appendix C)
\be \label{finbbmads04} B^I|\phi_{s'}\rangle = b^0_{s'} \MTIL^I
|\phi_{s'}\rangle + b^-_{s'}A_{s'-1}^I |\phi_{s'-1}\rangle +
b^+_{s'}\bar{\alpha}^I|\phi_{s'+1}\rangle\,, \ee
where the coefficients $b^\pm_{s'}$, $b^0_{s'}$ are given by
\beq  \label{finbbmads05} && b^-_{s'}  = f(E_0,{\bf h},s')\,,
\\[5pt]
\label{finbbmads06} && b^+_{s'} = f^*(E_0,{\bf h},s'+1) \,,
\\[5pt]
\label{finbbmads07} && b^0_{s'} = {\rm i}\frac{(h_1+1) h_2
(E_0-2)}{s'(s'+1)} \eeq
and a complex-valued function $f(E_0,{\bf h},s')$ is found to be
\be \label{finbbmads08} |f(E_0,{\bf h},s')|^2  \equiv
\frac{\left((h_1+1)^2 -s'{}^2\right)\left(s'{}^2
-h_2^2\right)\left((E_0-2)^2-s'{}^2\right)}{(2s'+1)s'{}^2}\,. \ee
The operators $\widetilde M^I$, $A_s^I$ that enter the operator
$B^I$ in \rf{finbbmads04} are defined by the relations
\beq && \label{sof05} \MTIL^I \equiv
\epsilon^{IJK}\alpha^J\bar\alpha^K \,,
\\
&& \label{sof12} A_s^I \equiv \alpha^I
-\frac{\alpha^2\bar{\alpha}^I}{2s+1}\,,\qquad \alpha^2\equiv
\alpha^I\alpha^I\,, \eeq
where $\epsilon^{IJK}$ ($\epsilon^{123}=1$)  is the Levi-Civita
tensor of $so(3)$. Note that the operators $M^{IJ}$
\rf{spiopemij1} and $\MTIL^I$ \rf{sof05} are related by the
formulas
\be\label{MTILMIJ} \MTIL^I =
\frac{1}{2}\epsilon^{IJK}M^{JK}\,,\quad\qquad M^{IJ}
=\epsilon^{IJK}\MTIL^K\,.\ee
Below, we present various relations for the operators $\MTIL^I$
and $A_s^I$ which are helpful in solving the defining equation for
the spin operator $B^I$ given in \rf{basequ0}:
\beq \label{sof13} && A_s^I A_{s-1}^J  -(I\leftrightarrow J) =0\,,
\\[5pt]
\label{sof14} && \bar\alpha^I \MTIL^J  - \bar\alpha^J \MTIL^I
\approx - \epsilon^{IJK} (\alpha\bar\alpha +2)\bar\alpha^K\,,
\\
\label{sof15} && \MTIL^I\bar\alpha^J - \MTIL^J\bar\alpha^I \approx
\epsilon^{IJK}\alpha\bar\alpha \bar\alpha^K \,,
\\[5pt]
\label{sof16} && A_{s}^I \MTIL^J - A_{s}^J \MTIL^I \approx
\epsilon^{IJK} sA_{s}^K \,,
\\
\label{sof17} &&  \MTIL^I A_{s}^J  -  \MTIL^J A_{s}^I \approx
-\epsilon^{IJK} (s+2) A_{s}^K \,,
\\[5pt]
\label{sof18} && A_s^I \bar{\alpha}^J - (I\leftrightarrow J)
=M^{IJ}\,,
\\
\label{sof19} && \bar{\alpha}^I A_s^J - (I\leftrightarrow J) =-
\frac{2s+3}{2s+1} M^{IJ}\,, \eeq
where we use the sign $\approx$ to indicate those relations that
are valid on the space of ket-vectors subject to the constraints
\rf{homcon1},\rf{tracon1}.

A few remarks are in order.

i) From formula  \rf{finbbmads08}, it is easy to see that if
$h_1>|h_2|$, then $E_0$ should satisfy the restriction $ E_0 \geq
2 + \max s'$. Taking into account that $\max s'= h_1$, we find a
restriction on $E_0$,
\be E_0 \geq h_1 +  2\,.\ee
The boundary values $E_0=h_1+2$ correspond to the massless fields
\rf{bose01xx}, while values $E_0 > h_1 + 2$ correspond to the
massive fields \rf{bose01}. Thus, our analysis involves massless
fields \rf{bose01xx} as a particular case and reproduces the
restriction on $E_0$ corresponding to the massive fields
\rf{bose01}. Moreover, we will demonstrate below that our results
involve the self-dual massive fields \rf{bose01x} as a particular
case.

ii) {}In formulas \rf{finbbmads05}-\rf{finbbmads08}, phase factors
of the coefficients $b_{s'}^\pm$
\rf{finbbmads05},\rf{finbbmads06}, which are determined by phase
factors of the functions $f(E_0,{\bf h},s')$, have not been fixed.
It is easy to demonstrate that making use of field redefinitions,
the phase factors of $b_{s'}^\pm$ could be normalized to be equal
to 1, i.e., all coefficients $b_{s'}^\pm$ could be chosen
real-valued and positive. However, for flexibility we do not fix
phase factors of the coefficients $b_{s'}^\pm$.

iii) We found a realization of the spin operator $B^I$ on the
vectors $|\phi_{s'}\rangle$ which depends on $s'$
\rf{finbbmads04}. In contrast to this, the realization of the spin
operator $M^{IJ}$ given in \rf{spiopemij1} does depend on $s'$,
i.e., the operator $M^{IJ}$ takes the same form for all vectors
$|\phi_{s'}\rangle$. Now we would like to describe a realization
of the operator $B^I$, which also does not depend on $s'$. To this
end, we introduce new creation and annihilation oscillators
$\zeta$, $\bar\zeta$:
\be\label{betaosc} [\bar\zeta,\zeta]=1\,,\qquad
\bar\zeta|0\rangle=0\ee
and build the new ket-vector
\be \label{bossec20} |\phi\rangle = \sum_{s'= |h_2|}^{h_1}
\frac{\zeta^{h_1 - s'}}{\sqrt{(h_1 - s')!}} |\phi_{s'}\rangle\,,
\ee
where the normalization factors in expansion \rf{bossec20} are
chosen so as to keep the normalization used in the scalar product
\rf{scaprod01}. For this realization of the spin degrees of
freedom we obtain the following representation for the operator
$B^I$ on the vector $|\phi\rangle$ \rf{bossec20}:
\be\label{BIalt} B^I =  \hat f^0(E_0,{\bf h},\alpha\bar\alpha)
\MTIL^I +\frac{ \hat f(E_0,{\bf
h},\alpha\bar\alpha)}{\alpha\bar\alpha - \frac{1}{2}} k^I
\bar\zeta + \hat f^\dagger(E_0,{\bf
h},\alpha\bar\alpha+1)\zeta\bar\alpha^I\,, \ee
where the operator-valued functions $\hat f$, $\hat f^0$ are
defined by\footnote{
The operator $\alpha\bar\alpha$ appearing in denominators of
expressions \rf{BIalt}-\rf{fatf0} is well defined because this
operator acts on the vector $|\phi\rangle$ \rf{sof00ket} that does
not involve terms of degree zero in $\alpha^I$ (see
\rf{bose01x2}).}
\beq && |\hat f(E_0,{\bf h},\alpha\bar\alpha)|^2 \equiv
\frac{\left( h_1
+1+\alpha\bar\alpha\right)\left((\alpha\bar\alpha)^2
-h_2^2\right)\left( (E_0-2)^2 - (\alpha
\bar\alpha)^2\right)}{(2\alpha\bar\alpha
+1)(\alpha\bar\alpha)^2}\,,
\\[5pt]
\label{fatf0}&& \hat f^0(E_0,{\bf h},\alpha\bar\alpha) = {\rm
i}\frac{(h_1+1) h_2 (E_0-2)}{\alpha\bar\alpha(\alpha\bar\alpha +
1)} \eeq
and we use the convention $|\hat f|^2 \equiv \hat f\hat
f^\dagger$. The operator $k^I$ in \rf{BIalt} is defined by
\be\label{kIdefbos} k^I \equiv -\frac{1}{2}\alpha^2\bar\alpha^I
+\alpha^I(\alpha\bar\alpha + \frac{1}{2}) \ee
and this operator is a counterpart of the operator $A_s^I$ in
\rf{sof12} used in the representation for $B^I$ given in
\rf{finbbmads04}. The operator $k^I$ is similar to the conformal
boost operator and one readily verifies the commutator
$[k^I,k^J]=0$.

The representations for the operator $B^I$ given in
\rf{finbbmads04} and \rf{BIalt} are related as
\be\label{BIBI} B^I|\phi\rangle = \sum_{s'= |h_2|}^{h_1}
\frac{\zeta^{h_1 - s'}}{\sqrt{(h_1 - s')!}}\, B^I
|\phi_{s'}\rangle\,, \ee
where in the l.h.s. of \rf{BIBI} we should use the representation
given in \rf{BIalt}, while in the r.h.s. of \rf{BIBI} we should
use the representation given in \rf{finbbmads04}.

\newsection{Fermionic fields}

Before studying the concrete form of the spin operators $M^{IJ}$
and $B^I$ we should fix a field theoretical realization of the
spin degrees of freedom collected in $|\psi\rangle$. To discuss
field theoretical description of fermionic massive field, we could
use {\it complex-valued} tensor-spinor field $\psi$ that is
associated with the $so(4)$ algebra representation of weight ${\bf
h}$. However, we prefer to decompose such a field into traceless
totally symmetric tensor-spinor fields of the $so(3)$ algebra
$\psi^{I_1\ldots I_{s'}\alpha}$, $I=1,2,3$; $s'= |h_2|- \half,
|h_2|+\half, \ldots, h_1 - \half:$
\be \psi = \sum_{s'= |h_2| - \half}^{h_1- \frac{1}{2}} \oplus\,
\psi^{I_1\ldots I_{s'}\alpha}\,. \ee
As before, to avoid cumbersome expressions we use the creation and
annihilation oscillators $\alpha^I$ and $\bar{\alpha}^I$
\rf{bososc} and make use of the ket-vectors $|\psi_{s'}\rangle$
defined by
\be\label{genfun2} |\psi_{s'}\rangle \equiv \alpha^{I_1}\ldots
\alpha^{I_{s'}} \psi^{I_1\ldots I_{s'}\alpha}(x)|0\rangle\,. \ee
The spinor index $\alpha=1,2,3,4$ is to remind that we deal with
Dirac tensor-spinor fields $|\psi_{s'}\rangle$. Below, all spinor
indices are implicit. The ket-vector $|\psi_{s'}\rangle$ satisfies
the algebraic constraints
\beq \label{picon} && \Pi^\oplus |\psi_{s'} \rangle = |\psi_{s'}
\rangle\,,
\\
\label{homcon2}&& (\alpha\bar{\alpha}-s')|\psi_{s'} \rangle =0\,,
\\
\label{gamtra1}&& \gamma\bar\alpha|\psi_{s'} \rangle =0\,, \quad
\qquad\qquad \gamma\bar\alpha\equiv \gamma^I\bar\alpha^I\,,
\\
\label{tracon2}&& \bar{\alpha}^I\bar\alpha^I|\psi_{s'}
\rangle=0\,. \eeq
Constraint \rf{picon} is the standard constraint of the light-cone
formalism, which allows us to deal with physical degrees of
freedom of the Dirac tensor-spinor field (some details may be
found in Appendices A and B). Equation \rf{homcon2} tells us that
$|\psi_{s'}\rangle$ is a monomial of degree $s'$ in the oscillator
$\alpha^I$. Tracelessness of the tensor-spinor field
$\psi^{I_1\ldots I_{s'}\alpha}$ is reflected in \rf{tracon2}.
Thus, in the language of ket-vectors the massive field $\psi$ is
representable as
\be \label{sof00ketfer} |\psi\rangle = \sum_{s'=|h_2|-\half
}^{h_1-\half} \oplus\, |\psi_{s'}\rangle \ee
and the scalar product $\langle\chi||\psi\rangle$ that enters
light-cone action \rf{lcactfer} is defined to be
\be\label{fersec10}  \langle\chi| |\psi\rangle  \equiv
\sum_{s'=|h_2|-\half}^{h_1-\half}
\langle\chi_{s'}||\psi_{s'}\rangle \,.\ee
Realization of the spin operator $M^{IJ}$ on the space of
ket-vectors $|\psi_{s'}\rangle$ is given by
\beq\label{MIJfer} M^{IJ} =
\alpha^I\bar\alpha^J-\alpha^J\bar\alpha^I+\frac{1}{2}\gamma^{IJ}\,,
\qquad  \gamma^{IJ}\equiv\frac{1}{2}(
\gamma^I\gamma^J-\gamma^J\gamma^I)\,.\eeq
The action of the operator $B^I$ on the physical  fermionic fields
$|\psi_{s'}\rangle$ is then found to be
\be \label{ferfinbbmads04} B^I|\psi_{s'}\rangle = b^0_{s'} \MTIL^I
|\psi_{s'}\rangle + b^-_{s'}\AA_{s'-1}^I |\psi_{s'-1}\rangle +
b^+_{s'}\bar{\alpha}^I|\psi_{s'+1}\rangle\,. \ee
As before, the coefficients $b_{s'}^\pm$, $b_{s'}^0$ depend on
$E_0$, $h_1$, $h_2$, $s'$ and are given by
\beq  \label{ferfinbbmads05} && b^-_{s'}  = f(E_0,{\bf h},s')\,,
\\[5pt]
\label{ferfinbbmads06} && b^+_{s'} = f^*(E_0,{\bf h},s'+1) \,,
\\[5pt]
\label{ferfinbbmads07} && b^0_{s'} = {\rm i}\frac{(h_1+1) h_2
(E_0-2)}{(s' +\frac{1}{2})(s' +\frac{3}{2})}\,, \eeq
where a complex-valued function $f(E_0,{\bf h},s')$ is found to be
\be \label{ferfinbbmads08}\hspace{-0.6cm} |f(E_0,{\bf h},s')|^2
\equiv \frac{\left((h_1+1)^2 - (s'+\frac{1}{2})^2\right)\left(
(s'+\frac{1}{2})^2 -h_2^2\right)\left((E_0-2)^2 -
(s'+\frac{1}{2})^2\right)}{(2s'+1)s'(s'+1)}\,. \ee
The operators $\widetilde M^I$ and $\AA_{s}^I$ that enter the spin
operator $B^I$ in \rf{ferfinbbmads04} are given by
\beq \label{marel01} && \MTIL^I \equiv \epsilon^{IJK}
\alpha^J\bar\alpha^K  + \frac{\rm i}{2}\gamma^I\,,
\\
\label{marel02} && \AA_s^I \equiv \alpha^I -\frac{\alpha^2
\bar\alpha^I +\gamma\alpha \gamma^I}{2s+3}\,. \eeq
Here, we present various relations for the operators $\MTIL^I$ and
$\AA_s^I$, which are helpful in solving the defining equation for
the spin operator $B^I$ given in \rf{ferfinbbmads04}:
\beq  \label{marel05} && \AA_s^I \AA_{s-1}^J  -(I\leftrightarrow
J) =0\,,
\\[5pt]
\label{marel06} && \bar\alpha^I \MTIL^J  -(I\leftrightarrow J)
\approx -(\alpha\bar\alpha+\frac{5}{2})\epsilon^{IJK}
\bar\alpha^K\,,
\\
\label{marel07} &&  \MTIL^I \bar\alpha^J   -(I\leftrightarrow J)
\approx (\alpha\bar\alpha + \frac{1}{2})\epsilon^{IJK}
\bar\alpha^K\,,
\\[5pt]
\label{marel08} && \AA_s^I \MTIL^J  -(I\leftrightarrow J) \approx
(s+\frac{1}{2})\epsilon^{IJK} \AA_s^K\,,
\\
\label{marel09} &&  \MTIL^I \AA_s^J   -(I\leftrightarrow J)
\approx -(s + \frac{5}{2})\epsilon^{IJK} \AA_s^K\,,
\\[5pt]
\label{marel10} && \AA_s^I \bar{\alpha}^J - (I\leftrightarrow J)
\approx \frac{2(s+1)}{2s+3}M^{IJ}\,,
\\
\label{marel11} && \bar{\alpha}^I \AA_s^J - (I\leftrightarrow J)
\approx - \frac{2(s+2)}{2s+3} M^{IJ}\,, \eeq
where we use the sign $\approx$ to indicate those relations that
are valid on the space of ket-vectors subject to the constraints
\rf{picon}-\rf{tracon2}.

The above realization of the spin operator $B^I$
\rf{ferfinbbmads04} on the vectors $|\psi_{s'}\rangle$ depends on
$s'$. By analogy with bosonic fields this realization can be
rewritten in the form independent of $s'$. To this end, we use the
oscillators $\zeta$, $\bar\zeta$ \rf{betaosc} and build the
ket-vector
\be\label{fersec20} |\psi\rangle = \sum_{s'=
|h_2|-\frac{1}{2}}^{h_1-\frac{1}{2}} \frac{\zeta^{h_1-\frac{1}{2}
- s'}}{\sqrt{(h_1 -\frac{1}{2} - s')!}} |\psi_{s'}\rangle\,, \ee
where normalization factors in expansion \rf{fersec20} are chosen
so as to keep normalization of the scalar product used in
\rf{fersec10}. The realization of spin degrees of freedom given in
\rf{fersec20} leads to the following representation for the
operator $B^I$ on the vector $|\psi\rangle$ \rf{fersec20}:
\be B^I =  \hat f^0(E_0,{\bf h},\alpha\bar\alpha) \MTIL^I  +
\frac{ \hat f(E_0,{\bf h},\alpha\bar\alpha)}{\alpha\bar\alpha -
\frac{1}{2}} k^I \bar\zeta + \hat f^\dagger(E_0,{\bf h},\alpha
\bar\alpha+1)\zeta\bar\alpha^I\,, \ee
where operator-valued functions $\hat f^0$, $\hat f$ are defined
by
\beq && |\hat f(E_0,{\bf h},\alpha\bar\alpha)|^2 \equiv \frac{(
h_1 +\frac{3}{2}+\alpha\bar\alpha)\left((\alpha\bar\alpha +
\frac{1}{2})^2 -h_2^2\right)\left( (E_0-2)^2 - (\alpha \bar\alpha
+ \frac{1}{2})^2\right)}{(2\alpha\bar\alpha
+1)\alpha\bar\alpha(\alpha\bar\alpha + 1)}\,, \ \ \ \ \ \
\\[6pt]
&& \hat f^0(E_0,{\bf h},\alpha\bar\alpha) = {\rm i}\frac{(h_1+1)
h_2 (E_0-2)}{(\alpha\bar\alpha + \frac{1}{2})(\alpha\bar\alpha +
\frac{3}{2})}\,.\eeq
As before we use the convention $|\hat f|^2 \equiv \hat f\hat
f^\dagger$, while the operator $k^I$ is modified compared to that
in \rf{kIdefbos} as follows
\be k^I \equiv -\frac{1}{2}\alpha^2\bar\alpha^I
+\alpha^I(\alpha\bar\alpha + 1) +
\frac{1}{2}\gamma^{IJ}\alpha^J\,. \ee
This operator satisfies the commutator $[k^I,k^J]=0$.

\newsection{Various limits of mixed-symmetry massive fields}

In previous sections, we found the light-cone gauge actions
\rf{lcact},\rf{lcactfer} for the mixed-symmetry massive fields
\rf{bose01x2}. Note that these actions can also be used for
description of totally symmetric massive fields \rf{bose01x1}. All
that is required to obtain the actions for totally symmetric
fields is to set $h_2=0$ (for bosonic fields) and $h_2 = \pm 1/2$
(for fermionic fields) in the above actions for mixed-symmetry
fields. In various limits the actions \rf{lcact},\rf{lcactfer}
lead to actions for self-dual massive fields \rf{bose01x} and
massless fields \rf{bose01xx} which were discussed in detail in
\cite{rrm02226}. Another interesting case is the flat space limit.
In this case, our actions become the actions for massive fields
propagating in flat space. We discuss these limits in turn.

\subsection{Limit of massless fields in $AdS_5$}

To realize limit of massless fields \rf{bose01xx}, we take
\be\label{m0lim} E_0 \rightarrow h_1 +2\,,\qquad  h_1 > |h_2|\,.
\ee
It can be verified that this limit leads to appearance of the
respective bosonic and fermionic invariant subspaces in
$|\phi\rangle$ \rf{sof00ket} and $|\psi\rangle$ \rf{sof00ketfer}
and these invariant subspaces, denoted as
$\phi_{E_0=h_1+2}^{subsp}$ and $\psi_{E_0=h_1+2}^{subsp}$, are
given by highest terms in the respective expansions \rf{sof00ket}
and \rf{sof00ketfer}:
\beq && \phi_{E_0=h_1+2}^{subsp} = |\phi_{h_1}\rangle\,,
\quad\qquad\qquad \ \hbox{for bosonic fields},
\\[4pt]
&& \psi_{E_0=h_1+2}^{subsp} = |\psi_{h_1-\frac{1}{2}}\rangle\,,
\qquad\qquad \hbox{for fermionic fields}, \eeq
In other words, in the limit \rf{m0lim}, the vectors
$|\phi_{h_1}\rangle$ and $|\psi_{h_1 - 1/2}\rangle$ transform into
themselves under the action of the respective spin operators
$M^{IJ}$ \rf{spiopemij1},\rf{MIJfer} and $B^I$
\rf{finbbmads04},\rf{ferfinbbmads04} and these vectors describe
spin degrees of freedom of the respective bosonic and fermionic
massless fields \rf{bose01xx}. The realization of the spin
operator $M^{IJ}$ on $|\phi_{h_1}\rangle$ and $|\psi_{h_1 -
1/2}\rangle$ is given by \rf{spiopemij1} and \rf{MIJfer}
respectively. To find the realization of the spin operator $B^I$
on the vectors $|\phi_{h_1}\rangle$ and $|\psi_{h_1 -
1/2}\rangle$, we note that in limit \rf{m0lim}, the action of the
operator $B^I$ on $|\phi_{h_1}\rangle$ \rf{finbbmads04} and
$|\psi_{h_1 - 1/2}\rangle$ \rf{ferfinbbmads04} is governed by
terms proportional to $b_{h_1}^0$, $b_{h_1-1/2}^0$ respectively.
By taking limit \rf{m0lim} in \rf{finbbmads07},\rf{ferfinbbmads07}
we obtain
\be b_{h_1}^0={\rm i}h_2,\qquad  b_{h_1-\frac{1}{2}}^0={\rm
i}h_2\ee
and this leads to the following realization for the operator
$B^I$:
\be\label{BIm0} B^I = {\rm i}h_2 \MTIL^I\,, \ee
which is valid for both bosonic $|\phi_{h_1}\rangle$ and fermionic
$|\psi_{h_1 - 1/2}\rangle$ fields. The realization of the $AdS$
mass operator $A$ on the massless fields $|\phi_{h_1}\rangle$ and
$|\psi_{h_1 - 1/2}\rangle$ can then be obtained using formula
\rf{adsope}. To this end we should evaluate $B^z$. Making use of
the convention $\epsilon^{zij}=\epsilon^{ij}$, $\epsilon^{12} = -
\epsilon^{21}= 1$ and the respective formulas for the spin
operator $\MTIL^I$ given in \rf{sof05},\rf{marel01} we obtain the
following representation for $AdS$ mass operator \cite{rrm02226}:
\be A =-\frac{1}{2}m^{ij}m^{ij} - \frac{1}{4}\,,\qquad
m^{ij}\equiv M^{ij} -{\rm i}\epsilon^{ij}h_2\,,\ee
where the $so(2)$ spin operator $M^{ij}$ takes the form given in
\rf{spiopemij1} and \rf{MIJfer} for bosonic and fermionic fields
respectively.

We finish with comparison of the number of physical degrees of
freedom in $AdS$ and flat spaces. On the one hand the number of
real-valued physical degrees of freedom ($DoF$) for {\it
mixed-symmetry massless} field \rf{bose01xx}\footnote{
Massless fields \rf{bose01xx} with $|h_2|> 1/2$ are referred to as
mixed-symmetry fields, while those with $h_2=0,\pm1/2$ are
referred to as totally symmetric fields.}
in $AdS$ space is equal to $2(2h_1+1)$ (see Refs.
\cite{rrm02226}), while the numbers of the $DoF$ for {\it totally
symmetric bosonic and fermionic massless} fields are equal to
$2h_1+1$ and $2(2h_1+1)$ respectively. On the other hand, it is
well known that the numbers of the $DoF$ for massless spin $h_1$
bosonic and fermionic fields\footnote{
In $5d$ flat space physical $DoF$ of the massless spin $h_1$ field
transform in irreps of the $so(3)$ algebra. Such a field has the
only label $h_1$ that is a weight of the $so(3)$ algebra irreps,
and there is no label similar to $h_2$. Therefore, all massless
fields in $5d$ flat space can be considered totally symmetric
fields.}
in flat space are equal to $2h_1+1$ and $2(2h_1+1)$ respectively.
This implies that:

i) the number of the $DoF$ for {\it mixed-symmetry massless
bosonic} field \rf{bose01xx} in $AdS$ space is twice that for the
{\it massless spin $h_1$ bosonic} field in flat space-time, while
the number of the $DoF$ for {\it mixed-symmetry massless
fermionic} field \rf{bose01xx} in $AdS$ space is equal to that for
{\it massless fermionic fields} in flat space-time\footnote{
We thank K.Alkalaev for raising the question concerning the
comparison of $DoF$ for {\it fermionic} fields in $AdS$ space and
those in flat space.};

ii) the number of the $DoF$ for totally symmetric massless field
is equal to that for massless field in flat space-time (this was
well known previously).

\subsection{Limit of self-dual massive fields in $AdS_5$}

The limit of self-dual massive fields \rf{bose01x} is realized by
taking
\be\label{selflim}  h_1 \rightarrow |h_2|\,. \ee
In view of $h_1 = |h_2|$ and formulas
\rf{sof00ket},\rf{sof00ketfer},  we see that physical degrees of
freedom are described by
\beq && |\phi\rangle\bigl|_{h_1= |h_2|} = |\phi_{h_1}\rangle\,,
\quad\qquad\qquad \ \hbox{for bosonic fields},
\\[3pt]
&& |\psi\rangle\bigl|_{h_1 =|h_2|} =
|\psi_{h_1-\frac{1}{2}}\rangle\,, \qquad\qquad \hbox{for fermionic
fields}. \eeq
The spin operators $M^{IJ}$ preserve their form given in
\rf{spiopemij1},\rf{MIJfer}, while the operator $B^I$ is reduced
to the operator $\MTIL^I$ and is governed by terms proportional to
$b_{h_1}^0$, $b_{h_1 - 1/2}^0$. By taking limit \rf{selflim} in
\rf{finbbmads07},\rf{ferfinbbmads07}, we obtain the respective
coefficients for the bosonic and fermionic fields
\be b_{h_1}^0 ={\rm i} (E_0-2) sign\, h_2, \qquad
b_{h_1-\frac{1}{2}}^0={\rm i} (E_0-2) sign\, h_2\,, \ee
where $sign\ h = +1(-1)$ for $h>0$($h<0$). These expressions lead
to the following representation for the spin operator $B^I$:
\be B^I = {\rm i}  (E_0-2) sign\, h_2\, \MTIL^I\,. \ee
Evaluating $B^z$ and making use of formula \rf{adsope}, we obtain
the $AdS$ mass operator \cite{rrm02226}:
\be  A =-\frac{1}{2}m^{ij}m^{ij} - \frac{1}{4}\,,\qquad
m^{ij}\equiv M^{ij} -{\rm i}\epsilon^{ij}(E_0-2)\, sign\,h_2\,,\ee
where the operator $M^{ij}$ is given in \rf{spiopemij1} for
bosonic fields and in \rf{MIJfer} for fermionic fields.

\subsection{Flat space limit}

To realize the flat space limit we introduce a new coordinate
$\phi$ defined by
\be\label{zphi} z = R e^{\phi/R}\,,\ee
and take
\be\label{flalim02} R \rightarrow \infty\,,\qquad \phi,\,\,
x^\pm,\,\, x^i -\hbox{ fixed}\,,\ee
where $x^\pm$, $x^i$ are four isometric  coordinates  along the
boundary directions (see \rf{Poicor}). In this limit the
light-cone actions for $AdS$ massive fields
\rf{lcact},\rf{lcactfer} become the actions for massive fields
propagating in flat space-time:
\beq && \label{lcactfl} S_{l.c.} =\frac{1}{2}\int d^5 x \langle
\phi|\bigl(\Box - m^2\bigr)|\phi\rangle\,, \qquad\quad\ \
\hbox{for bosonic fields},
\\[5pt]
\label{lcactferfl} && S_{l.c.}= \int d^5 x\, \langle \psi|
\frac{\rm i}{2\partial^+}(\Box - m^2)|\psi\rangle\,, \qquad
\hbox{for fermionic fields}. \eeq
Note that in this limit the number of physical degrees of freedom
does not change, i.e., we can use the fields $|\phi\rangle$ in
\rf{scaprod01} and $|\psi\rangle$ in \rf{sof00ketfer} for
description of the respective bosonic and fermionic massive fields
in flat space-time. In other words {\it arbitrary spin massive
mixed-symmetry fields in $AdS$ space and those in flat space have
the same number of physical degrees of freedom}. Let us comment on
how the actions in flat space \rf{lcactfl},\rf{lcactferfl} can be
obtained from those in $AdS$ space \rf{lcact} \rf{lcactfer}.

Relations between various quantities that we are using for the
description of $AdS$ and flat space massive fields can be obtained
by exploiting the two relations
\beq\label{E0Rm} && E_0 \rightarrow R m\,,
\\[3pt]
&& \label{MIJbfMIJ} M^{IJ} = \bfM^{IJ}\,,\eeq
where $\bfM^{IJ}$ stands for the $so(d-2)$ algebra spin operator
of the massive field in flat space. Here and below, we use
boldface letters for the generators of the Poincar\'e algebra and
spin operators of massive fields in flat space. Relation \rf{E0Rm}
is well known from the study of various massive fields with
particular values of spin and it can also be easily obtained from
the general results of
Refs.\cite{Metsaev:2003cu}\footnote{
Formulas (5.62),(5.63) of Refs.\cite{Metsaev:2003cu} which relate
$E_0$ and the mass parameter $m$ for massive fields of arbitrary
spin are given for the normalization of $AdS$ radius $R=1$. In
order to restore the dependence on the radius $R$ space, one needs
to make the rescaling $m \rightarrow Rm$ in those formulas.}.
Relation \rf{MIJbfMIJ} reflects the fact that the spin operator of
the $so(d-2)$ algebra in $AdS$ and flat space does not depend on
$R$ at all. Making use of \rf{E0Rm},\rf{MIJbfMIJ} and of the
equation for the $AdS$ mass operator \rf{adsope} and the spin
operator $B^I$ \rf{basequ0}, it is easy to establish the following
relations in limit as $R\rightarrow \infty$:
\beq\label{flalim20} && A \rightarrow R^2m^2 \,,
\\
\label{flalim21}  && B^I \rightarrow  R m \bfM^I \,,\eeq
where the flat space spin operator $\bfM^I$ satisfies the
commutators
\be [\bfM^I, \bfM^{JK}] = \delta^{IJ}\bfM^K - \delta^{IK}\bfM^J
\,, \quad [\bfM^I,\bfM^J ] = \bfM^{IJ}\,. \ee
{}From \rf{flalim21}, we see that the spin operator $B^I$ is an
$AdS$ counterpart of the flat space spin operator $\bfM^I$. We
note that $\bfM^I$ and $\bfM^{IJ}$ form the commutators of the
$so(d-1)$ algebra and satisfy the following hermitian conjugation
rules $\bfM^{IJ\dagger }= - \bfM^{IJ}$, $ \bfM^{I\dagger }=
\bfM^I$. Plugging formulas \rf{zphi},\rf{flalim20} in \rf{lcact}
\rf{lcactfer}, we see that actions \rf{lcact} \rf{lcactfer} for
massive fields in AdS space are indeed reduced to those in flat
space \rf{lcactfl},\rf{lcactferfl}. We finish with the description
of interrelations between the relativistic symmetries in $AdS$
space and those in flat space.

In the limit \rf{flalim02} the generators of the $AdS$ algebra
given in \rf{pi}-\rf{km} tend to those of the Poincar\'e algebra
as
\be \label{Pmlim} P^- \rightarrow \bfP^-\,, \qquad   J^{-i}
\rightarrow \bfJ^{-i} \,,  \qquad  J^{+-} \rightarrow
\bfJ^{+-}\,,\ee
\beq\label{DPz} && D \rightarrow R \bfP^z \,,
\\
&& K^i \rightarrow -\frac{1}{2}R^2 \bfP^i + R\bfJ^{iz} \,,
\\
&& K^+ \rightarrow -\frac{1}{2}R^2 \bfP^+ + R\bfJ^{+z} \,,
\\
\label{Jmz} && K^- \rightarrow -\frac{1}{2}R^2 \bfP^- + R
\bfJ^{-z} \,. \eeq
The remaining generators of the $AdS$ algebra
\rf{pi},\rf{J+i},\rf{Jij} coincide with those of the Poincar\'e
algebra:
\be\label{flalim30} P^i=\bfP^i\,,\qquad P^+=\bfP^+\,,\qquad J^{+i}
= \bfJ^{+i}\,, \qquad J^{ij} = \bfJ^{ij}\,. \ee

To make our presentation self-contained, we write the
representation for generators of the Poincar\'e algebra which we
used in establishing the relations in \rf{Pmlim}-\rf{flalim30}:

\beq \label{bfPI} && \bfP^I=\partial^I\,,\qquad
\bfP^+=\partial^+\,,\qquad \bfP^-=\frac{-\partial^I\partial^I
+m^2}{2\partial^+}\,,
\\
&& \bfJ^{+I}= x^+\partial^I - x^I\partial^+\,,
\\
&& \bfJ^{+-}=  x^+ \bfP^- - x^-\partial^+\,,
\\
&& \bfJ^{IJ}=x^I\partial^J- x^J\partial^I + \bfM^{IJ}\,,
\\
\label{bfJmI}&& \bfJ^{-I}=x^-\partial^I - x^I \bfP^- +
\frac{1}{\partial^+}(\bfM^{IJ}\partial^J + m\bfM^I)\,. \eeq
In \rf{bfPI}-\rf{bfJmI}, the coordinates $x^I$ stand for $x^i$,
$\phi$. Accordingly, the derivatives $\partial_I$ stand for
$\partial_i =\partial/\partial x^i$, $\partial_\phi
=\partial/\partial \phi$. In \rf{DPz}-\rf{Jmz}, we use the
identifications $\bfP^z=\bfP^\phi$, $\bfJ^{\pm z}=\bfJ^{\pm
\phi}$.

\newsection{Conclusions}

The results presented here should have a number of interesting
applications and generalizations, some of which are: i) In this
paper, we develop a light-cone formulation for massive fields
propagating in $AdS_5$. It would be interesting to extend this
formulation to the study of massive arbitrary spin fields
propagating in $AdS_5\times S^5$; ii) AdS/CFT correspondence for
various $AdS_d$ massive fields and the corresponding boundary
operators was studied in \cite{Witten:1998qj, Henningson:1998cd,
pol,l'Yi:1998eu,Volovich:1998tj, Koshelev:1998tu} (see
\cite{D'Hoker:2002aw} for a review). AdS/CFT correspondence for
massless and self-dual massive fields in $AdS_5$ and the
corresponding boundary conformal operators has been demonstrated
at the level of two point functions in \cite{rrm02226} (see also
\cite{Germani:2004jf} for some related interesting discussion). It
would be interesting to extend the results of this paper to the
study of AdS/CFT correspondence for mixed-symmetry massive $AdS_5$
fields along the lines of
Refs.\cite{Dobrev:1998md,Vasiliev:2001zy,Bianchi:2003wx}.

In this paper, we studied the classical fields propagating in
$AdS_5$ space-time. It would be interesting to extend the results
of this paper to the case of quantized fields, to study the
space-time correlation functions of such fields, and to clarify
the relation between the positivity of the energy and the
analyticity properties of the correlation functions. We hope to
return to these problems in future publications.

{\bf Acknowledgments}. We thank A. Semikhatov for useful comments.
This work was supported by the INTAS project 03-51-6346, by the
RFBR Grant No.02-02-17067, and RFBR Grant for Leading Scientific
Schools, Grant No. 1578-2003-2.

\setcounter{section}{0} \setcounter{subsection}{0}

\appendix{Notation and conventions}

We use the Poincar\'e parametrization of $AdS_5$ space in which

\be\label{Poicor} ds^2= \frac{R^2}{z^2}(- dx^0 dx^0 + dx^idx^i +
dz dz + dx^4 dx^4)\,,\ee
where $R$ is the radius of $AdS_5$ space, and we use indices
\be i,j,k =1, 2\,; \qquad I,J,K=1,2,3\ee
and often identify the radial coordinate $z$ as
\be z = x^3 \,.\ee
The light-cone coordinates in the $\pm$ directions are defined as
\be x^\pm =  \frac{1}{\sqrt{2}} (x^4 \pm x^0)\ee
and $x^+$ is taken to be the light-cone time. We adopt the
conventions
\be \partial^I=\partial_I\equiv\partial/\partial x^I,\quad
\partial^\pm=\partial_\mp \equiv \partial/\partial x^\mp\,.\ee
The scalar product of the $so(3)$ algebra vectors $X^I=(X^i,X^z)$,
$Y^I=(Y^i,Y^z)$ is given by
\be X^IY^I=X^iY^i+X^zY^z\,,\qquad X^3 \equiv X^z, \quad Y^3 \equiv
Y^z\,.\ee
We use $4\times 4$ -Dirac $so(4,1)$ $\gamma$-matrices:
\be \{\gamma^a,\gamma^b\}=2\eta^{ab}\,, \qquad a,b=0,1,\ldots,
4\,,\ee
where $\eta^{ab}$ is the mostly positive flat metric tensor. The
$\gamma$-matrices are normalized as
\be \label{1app20}\gamma^0\gamma^1\gamma^2\gamma^3 \gamma^4 ={\rm
i}\,.\ee
In the light-cone frame, the $\gamma$-matrices are decomposed in
the standard way
\be \gamma^a=\gamma^+,\gamma^-,\gamma^I, \qquad  \gamma^\pm\equiv
\frac{1}{\sqrt{2}} (\gamma^4 \pm \gamma^0)\,.\ee
The four component Dirac spinor field of the $so(4,1)$ algebra
$\psi_{Dirac}$ can be decomposed as
\be\label{ferlcspl} \psi_{Dirac} = \psi^\oplus
+\psi^\ominus\,,\qquad
\psi^\oplus \equiv \Pi^\oplus \psi_{Dirac}\,,\qquad
\psi^\ominus \equiv \Pi^\ominus \psi_{Dirac}\,,\ee
where the standard projectors $\Pi^\oplus$, $\Pi^\ominus$ defined
by

\be \Pi^\oplus \equiv \frac{1}{2}\gamma^-\gamma^+\,,\qquad
\Pi^\ominus \equiv \frac{1}{2}\gamma^+\gamma^-\,, \ee
satisfy the relations
\be\label{1app26} \Pi^\oplus \Pi^\oplus = \Pi^\oplus\,, \qquad
\Pi^\ominus \Pi^\ominus =  \Pi^\ominus\,,\qquad \Pi^\ominus
\Pi^\oplus = 0\,,\ee
\be \Pi^\oplus +\Pi^\ominus =1\,. \ee
In the light-cone formalism, the field $\psi^\oplus$ turns out to
be physical field, while the field $\psi^\ominus$ is a
non-physical field (for more details concerning the spin one-half
Dirac field, see Appendix B). Therefore, the light-cone action is
formulated entirely in terms of the field $\psi^\oplus$.

It turns out that the relations for $\gamma$-matrices are
simplified on the space of physical fields $\psi^\oplus$. This is
to say that normalization \rf{1app20} leads to the relation
\be\label{1app28} \gamma^{IJ} = {\rm i}\epsilon^{IJK}\gamma^K
(\Pi^\oplus -\Pi^\ominus)\,,\ee
where $\epsilon^{IJK}$ is Levi-Civita symbol subject to the
normalization $\epsilon^{123}=1$. Applying relation \rf{1app28} to
$\psi^\oplus$ and using the first formula in \rf{1app26} we obtain
the simplified formula
\be \gamma^{IJ} \psi^\oplus  = {\rm i}\epsilon^{IJK}\gamma^K
\psi^\oplus\,.\ee
This formula implies that the $4\times 4$ $\gamma^I$-matrices
being restricted  to the space of physical fields  $\psi^\oplus$
satisfy the same relations as the standard $2\times 2$ Pauli
matrices. This property of the $4\times 4$ $\gamma^I$-matrices is
used throughout this paper. Note that in main body of the paper we
use the simplified notation $\psi$ for the physical field
$\psi^\oplus$ (i.e., we drop the superscript $\oplus$).

 \setcounter{subsection}{0}

\appendix{Derivation of light-cone action for spin one-half Dirac field}

Here we  discuss the transformation of the Lorentz covariant
action for the spin one-half Dirac field to the light-cone action
given in \rf{lcactfer},\rf{lcactfer}. The standard Lorentz
covariant action for the spin one-half Dirac field in $AdS_d$
\be\label{Sonehal01} S=  - \int d^dx\, e \bar{\Psi}(\gamma^\mu
D_\mu + m)\Psi\,, \qquad \bar{\Psi} \equiv \Psi^\dagger {\rm
i}\gamma^0 \,, \ee
considered in Poincar\'e coordinates \rf{Poicor}, takes the form
\be S = - \int d^dx\, \Bigl(\frac{z}{R}\Bigr)^{-d}
\bar{\Psi}\Bigl(\frac{z}{R}\gamma^a
\partial_{x^a} + \frac{1-d}{2R}\gamma^z + m \Bigr)\Psi\,. \ee
Introducing the canonically normalized field $\psi$,
\be \Psi = \Bigl(\frac{z}{R}\Bigr)^{\frac{d-1}{2}}\psi \ee
we obtain
\be\label{Sfer}  S=- \int \, d^dx\,\,  \bar{\psi} \Bigl(\gamma^a
\partial_a + \frac{Rm}{z}\Bigr)\psi\,. \ee
Making use of the light-cone splitting for the Dirac spinor
\be \psi^\oplus = q \Pi^\oplus \psi\,,\qquad \psi^\ominus =
q\Pi^\ominus \psi\,,\qquad q^2 = \sqrt{2}\,,\ee
where compared to \rf{ferlcspl} we use an extra normalization
factor $q$, we obtain the following light-cone representation for
the action \rf{Sfer}:
\beq \label{Sferlcspl} S  &= & \frac{\rm i}{2}\int\,d^dx\, \Bigl(
2\psi^{\oplus\dagger}\partial^-\psi^\oplus
-2\psi^{\ominus\dagger}\partial^+\psi^\ominus
\nonumber\\
&- &\psi^{\oplus\dagger}(\gamma^I\partial^I -
\frac{Rm}{z})\gamma^-\psi^\ominus
+\psi^{\ominus\dagger}(\gamma^I\partial^I - \frac{Rm}{z})\gamma^+
\psi^\oplus \Bigr)\,. \eeq
This action leads to the equation for the field $\psi^\ominus$,
\be \partial^+ \psi^\ominus =\frac{1}{2}(\gamma^I\partial^I -
\frac{Rm}{z})\gamma^+ \psi^\oplus\,, \ee
which tells us that $\psi^\ominus$ is not a dynamical field and
can be expressed in terms of the physical field $\psi^\oplus$.
Thus, by inserting the solution for $\psi^\ominus$ into action
\rf{Sferlcspl}, we obtain the light-cone action for the physical
field $\psi^\oplus$:
\be \label{Sonehal02} S_{l.c.} = \int\,d^dx\,\,
\psi^{\oplus\dagger}\frac{\rm i}{2\partial^+}\Bigl(\Box -
\frac{1}{z^2}\left((Rm)^2 + Rm\gamma^z\right)\Bigr)\psi^\oplus\,.
\ee
Comparison of \rf{Sonehal02} and \rf{lcactfer} leads to the $AdS$
mass operator given in \rf{Afer}.

 \setcounter{subsection}{0}

\appendix{Derivation of expression for spin operator $B^I$.}

Here, we describe the method of solving equations for the operator
$B^I$ given in \rf{basequ0}-\rf{BIunicon}. Unitarity equation
\rf{BIunicon} and $so(3)$ covariance equation \rf{bId2tra} are
easiest to treat. The main problem is to solve Eq.\rf{basequ0},
which is nonlinear in the spin operator $M^{IJ}$. Fortunately, it
turns out that Eq.\rf{basequ0} can be reduced to the analysis of
commutation relations for spin operators of the $so(4)$ algebra.
We demonstrate our method for the bosonic fields. Extension to the
fermionic fields is straightforward. Our method consists of the
following steps:

i) First, we start with the analysis of Eq.\rf{basequ0}. Our basic
observation is that a solution of Eq.\rf{basequ0} can be presented
as
\be\label{MIMIsol} B^I = y M^I - M^{IJ}M^J\,, \qquad y\equiv E_0-1\,,\ee
provided a new spin operator $M^I$  that transforms in the vector
representation with respect to $so(3)$ transformations generated
by $M^{IJ}$ satisfies the commutation relation
\be\label{bbm01} [ M^I, M^J] = M^{IJ}\,,\ee
and the constraint
\beq \label{genrep12} && M^IM^I -\frac{1}{2}M^{IJ}M^{IJ} +\langle
Q_{so(4)}\rangle =0\,,
\\
&& -\langle Q_{so(4)}\rangle \equiv h_1 (h_1 +2) +h_2^2\,, \eeq
where $\langle Q_{so(4)}\rangle$ is an eigenvalue of the second
order Casimir operator for irreps of the $so(4)$ algebra labelled
by ${\bf h}=(h_1,h_2)$. Because the operator $M^I$ transforms in
the vector representation of the $so(3)$ algebra transformation
generated by the spin operator $M^{IJ}$, the operators $M^I$ and
$M^{IJ}$ form commutators of the $so(4)$ algebra. Thus, we reduced
Eq.\rf{basequ0} to the problem of solving
Eqs.\rf{bbm01},\rf{genrep12}. Compared to non-linear
Eq.\rf{basequ0}, equations \rf{bbm01},\rf{genrep12} are easier to
solve because they are the standard equations for the spin
operator of the $so(4)$ algebra. We begin with the analysis of
Eq.\rf{bbm01}. To this end, we write the most general $so(3)$
covariant expression for the action of the operator $M^I$ on the
ket-vectors $|\phi_{s'}\rangle$:
\be \label{bbm05} M^I|\phi_{s'}\rangle = m^0_{s'} \MTIL^I
|\phi_{s'}\rangle + m^-_{s'}A_{s'-1}^I |\phi_{s'-1}\rangle +
m^+_{s'}\bar{\alpha}^I|\phi_{s'+1}\rangle\,, \ee
where the coefficients $m^0_{s'}$, $m^\pm_{s'}$ are still to be
fixed. The representation for $M^I$ given in \rf{bbm05} is fixed
simply by the requirement that action of $M^I$ on
$|\phi_{s'}\rangle$ should respect the constraints given in
\rf{homcon1},\rf{tracon1}. Equation \rf{bbm05} should be
supplemented by the `initial' conditions
\be\label{inicon} m_{|h_2|}^- = 0\,,\qquad m_{h_1}^+ =0\ee
which express the fact that the action of the operator $M^I$ is
defined on those vectors $|\phi_{s'}\rangle$ whose spin $s'$ takes
values in the domain $s'=|h_2|,\ldots,h_1$. Now we demonstrate
that the coefficients $m^0_{s'}$, $m^\pm_{s'}$ can be found by
analyzing Eq.\rf{bbm01} and `initial' conditions \rf{inicon}. To
this end, we write the action of two spin operators $M^I$ on
$|\phi_{s'}\rangle$:
\beq \label{bbm06} M^I M^J|\phi_{s'}\rangle &=& \Bigl(
m_{s'}^0{}^2 \MTIL^I \MTIL^J + m^-_{s'} m^+_{s'-1}
A_{s'-1}^I\bar\alpha^J+ m^-_{s'+1} m^+_{s'}\bar\alpha^I A_{s'}^J
\Bigr)|\phi_{s'}\rangle \nonumber
\\
&+& m^-_{s'} m^-_{s'-1} A_{s'-1}^I A_{s'-2}^J |\phi_{s'-2}\rangle
\nonumber\\
&+& \Bigl( m^0_{s'} m^-_{s'} \MTIL^I  A_{s'-1}^J + m^-_{s'}
m^0_{s'-1} A_{s'-1}^I \MTIL^J \Bigr)|\phi_{s'-1}\rangle
\nonumber\\
&+ &  \Bigl( m^0_{s'} m^+_{s'} \MTIL^I\bar{\alpha}^J + m^+_{s'}
m^0_{s'+1} \bar{\alpha}^I \MTIL^J \Bigr)|\phi_{s'+1}\rangle
\nonumber\\
&+& m^+_{s'} m^+_{s'+1}\bar\alpha^I\bar\alpha^J
|\phi_{s'+2}\rangle\,. \eeq
From this formula and \rf{sof13}-\rf{sof19}, we find that
commutator \rf{bbm01} gives the equations
\beq  \label{bbm10} && s' m^0_{s'} - (s'+2) m^0_{s'+1}=0\,,
\\
\label{bbm11} &&  - m_{s'}^0{}^2  + m^-_{s'} m^+_{s'-1} -
\frac{2s'+3}{2s'+1} m^-_{s'+1} m^+_{s'}   = 1\,. \eeq
Equations \rf{bbm10},\rf{bbm11} and \rf{inicon} lead to the
desired relations
\beq  \label{bbm22} && m^-_{s'} m^+_{s'-1}  =  \frac{((h_1+1)^2
-s'{}^2)(s'{}^2 -h_2^2)}{(2s'+1)s'{}^2}\,,
\\
\label{bbm23} && m_{s'}^0 = {\rm i} \frac{(h_1+1)h_2}{s'(s'+1)}\,.
\eeq
Making use of \rf{bbm22},\rf{bbm23} and the formulas
\beq \label{MIMI} &&  M^IM^I |\phi_{s'}\rangle = \Bigl(
h_1(h_1+2)+h_2^2 -s'(s'+1) \Bigr) |\phi_{s'}\rangle\,,
\\[3pt]
&& -\frac{1}{2} M^{IJ}M^{IJ}|\phi_{s'}\rangle  =
s'(s'+1)|\phi_{s'}\rangle \eeq
we check that Eq.\rf{genrep12} is satisfied automatically, i.e.,
Eq.\rf{genrep12} does not impose additional constraints on the
coefficients $m^0_{s'}$, $m^\pm_{s'}$. Helpful relations in
deriving \rf{MIMI} are
\beq && A_{s+1}^I A_s^I|\phi_s\rangle=0\,,
\\
&& \label{AIalI} A_{s-1}^I\bar\alpha^I |\phi_s\rangle =
s|\phi_s\rangle\,,
\\
\label{alIAI} && \bar\alpha^I A_s^I|\phi_s\rangle = \frac{(2s +
3)(s+1)}{2s + 1}|\phi_s\rangle\,. \eeq

ii) Second, we analyze the requirement that the spin operator
$B^I$ be hermitian with respect to the scalar product defined in
\rf{scaprod01}. Making use of formulas \rf{MIMIsol},\rf{bbm05}, we
find the following representation for the operator $B^I$:
\be \label{finbbmads04app} B^I |\phi_{s'}\rangle = b^0_{s'}
\MTIL^I |\phi_{s'}\rangle + b^-_{s'}A_{s'-1}^I |\phi_{s'-1}\rangle
+ b^+_{s'}\bar{\alpha}^I|\phi_{s'+1}\rangle\,, \ee
where coefficients $b^0_{s'}$,  $b^\pm_{s'}$ are expressible in
terms of $m^0_{s'}$,  $m^\pm_{s'}$ as
\beq\label{bsmmsm}  && b_{s'}^- = (y - s' - 1) m_{s'}^-\,,
\\
\label{bspmsp} && b_{s'}^+ = (y + s')m_{s'}^+\,,
\\
\label{bszmsz} && b_{s'}^0 = (y - 1) m_{s'}^0 \,. \eeq
These relations can be obtained from \rf{MIMIsol},\rf{bbm05} by
using the formulas
\beq \label{sof20} && M^{IJ}A_s^J|\phi_s\rangle = (s +
2)A_s^I|\phi_s\rangle\,,
\\[5pt]
\label{sof21} && M^{IJ}\bar\alpha^J|\phi_s\rangle =
(1-s)\bar\alpha^I|\phi_s\rangle\,,
\\[5pt]
\label{sof22} && M^{IJ}\MTIL^J = \MTIL^I\,. \eeq
The expression for $y$ in \rf{MIMIsol} and formulas
\rf{bbm23},\rf{bszmsz} lead to $b_{s'}^0$ given in
\rf{finbbmads07}.

Making use of the representation for the operator $B^I$ given in
\rf{finbbmads04app}, we find that the requirement for the operator
$B^I$ to be hermitian with respect to the scalar product in
\rf{scaprod01}
\be \langle \varphi||B^I\phi\rangle = \langle
B^I\varphi||\phi\rangle \ee
leads to the relations
\be \label{finbbmads09} b^+_{s'} = b^{-*}_{s'+1}\,,\qquad b^0_{s'}
= - b^{0*}_{s'}\,. \ee
Note that the coefficient $b_{s'}^0$ (see \rf{bbm23},\rf{bszmsz})
satisfies unitarity requirement \rf{finbbmads09} automatically.
Now we are ready to demonstrate that equations \rf{bbm22},
relations \rf{bsmmsm},\rf{bspmsp}, and unitarity requirement
\rf{finbbmads09} allow us to fix the coefficients $b^\pm_{s'}$.
Indeed, multiplying $b_{s'}^-$ by $b_{s'-1}^+$ and using formulas
\rf{bbm22},\rf{bsmmsm},\rf{bspmsp} we obtain
\be  b_{s'}^-b_{s'-1}^+ =  \frac{((h_1+1)^2 -s'{}^2)(s'{}^2
-h_2^2)}{(2s'+1)s'{}^2} ((E_0-2)^2 -s'{}^2)\,. \ee
{}From this formula and the first relation in \rf{finbbmads09}, we
obtain the solution for the coefficients $ b_{s'}^\pm $ given in
\rf{finbbmads05},\rf{finbbmads06}.

\end{document}